\documentclass[10pt, final, journal, letterpaper]{IEEEtran}
\pdfoutput=1
\usepackage{cite}
\usepackage{amsmath,amssymb,amsfonts}
\usepackage{algorithmic}
\usepackage{graphicx}
\usepackage{textcomp}

\usepackage[utf8]{inputenc} 
\usepackage[T1]{fontenc}    
\usepackage{hyperref}       
\usepackage{url}            
\usepackage{booktabs}       
\usepackage{nicefrac}       
\usepackage{microtype}
\usepackage{multirow}
\usepackage{makecell}
\usepackage{diagbox}
\usepackage[export]{adjustbox}
\usepackage{arydshln}
\usepackage{afterpage}
\usepackage[online,flushleft]{threeparttable}
\usepackage[multiple]{footmisc}
\usepackage[svgnames]{xcolor}
\def\@fnsymbol#1{\ensuremath{\ifcase#1\or *\or \dagger\or \ddagger\or
   \mathsection\or \mathparagraph\or \|\or **\or \dagger\dagger
   \or \ddagger\ddagger \else\@ctrerr\fi}}
   
\newcommand{\ssymbol}[1]{$^{\@fnsymbol{#1}}$}  
\newcommand{\REV}[1]{\textcolor{Black}{#1}}
\newcommand{\REVV}[1]{\textcolor{Black}{#1}}

\title{Deep Learning-Based Regression and Classification for Automatic Landmark Localization in Medical Images}
\author{Julia M. H. Noothout, Bob D. de Vos, Jelmer M. Wolterink, Elbrich M. Postma, Paul A. M. Smeets, \\Richard A. P. Takx, Tim Leiner, Max A. Viergever and Ivana I\v{s}gum%
	\thanks{Copyright (c) 2019 IEEE. Personal use of this material is permitted. However, permission to use this material for any other purposes must be obtained from the IEEE by sending a request to pubs-permissions@ieee.org. This work was supported by  the Dutch  Technology  Foundation  with  contribution  by Philips Healthcare under Project P15-26.}
	\thanks{Julia M. H. Noothout, Bob D. de Vos, and Ivana I\v{s}gum are with the Image Sciences Institute, University Medical \mbox{Center} Utrecht, Utrecht, The Netherlands, and with the Amsterdam UMC, University of Amsterdam, Biomedical Engineering and Physics, Amsterdam, The Netherlands \mbox{E-mail: jmhnoothout@gmail.com}}%
	\thanks{Jelmer M. Wolterink is with the University of Twente, Faculty of Electrical Engineering, Mathematics and Computer Science, Enschede, The Netherlands}%
	\thanks{Richard A.P. Takx and Tim Leiner are with the Department of Radiology, University Medical Center Utrecht, Utrecht, The Netherlands.}%
	\thanks{Elbrich M. Postma and Paul A.M. Smeets are with the Division of Human Nutrition and Health, Wageningen University, Wageningen, The Netherlands.}%
	\thanks{Max A. Viergever is with the Image Sciences Institute, University Medical \mbox{Center} Utrecht, Utrecht, The Netherlands}}

\begin{document}
    \maketitle
	\begin{abstract}
        In this study, we propose a fast and accurate method to automatically localize anatomical landmarks in medical images. \REVV{We employ a global-to-local localization approach using fully convolutional neural networks (FCNNs). First, a global FCNN localizes multiple landmarks through the analysis of image patches, performing regression and classification simultaneously. In regression, displacement vectors pointing from the center of image patches towards landmark locations are determined. In classification, presence of landmarks of interest in the patch is established. Global landmark locations are obtained by averaging the predicted displacement vectors, where the contribution of each displacement vector is weighted by the posterior classification probability of the patch that it is pointing from. Subsequently, for each landmark localized with global localization, local analysis is performed. Specialized FCNNs refine the global landmark locations by analyzing local sub-images in a similar manner, i.e. by performing regression and classification simultaneously and combining the results.} Evaluation was performed through localization of 8 anatomical landmarks in CCTA scans, 2 landmarks in olfactory MR scans, and 19 landmarks in cephalometric X-rays. We demonstrate that the method performs similarly to a second observer and is able \REV{to localize landmarks in a diverse set of medical images}, differing in image modality, image dimensionality, and anatomical coverage. 
	\end{abstract}

	
	\section{Introduction}
	\IEEEPARstart{I}{dentification} of anatomical reference points and landmarks is a prerequisite for numerous medical image analysis tasks\cite{rohr2001landmark}. These include image registration \cite{miao2012automatic, murphy2011semi, wang2018fast, han_robust_2014, alam2018medical}, initialization of segmentation methods \cite{oktay2017stratified}, and computation of clinical measurements for patient diagnosis and treatment planning \cite{wang2016benchmark, al2018automatic, torosdagli2018deep, kasel2013standardized, ionasec2008dynamic, zhengautomatic2010}. While manual identification of anatomical landmarks might be trivial, it is often tedious and cumbersome \cite{wang2016benchmark, elattar2016automatic}. Fast and accurate automatic landmark localization methods can replace manual identification and may be especially helpful when precise localization of multiple landmarks is required.
	
	Several application-specific automatic landmark localization methods have been proposed previously, such as methods combining segmentation of specific structures containing the landmarks and subsequent local rule-based analysis of those structures \cite{wachter2010patient, zheng2012automatic, elattar2016automatic}. More generic localization methods employ either multi-atlas image registration \cite{alam2018medical, isgum2009multi} or machine learning. In multi-atlas image registration, multiple atlas images with annotated landmarks are registered to the image of interest. Subsequently, a voting scheme determines the location of landmarks. Such approaches are accurate and robust to limited diversity in the anatomy and image acquisition, but they are typically time-consuming \cite{alam2018medical, isgum2009multi}. Machine learning provides a faster and more robust alternative. 
	
	Conventional machine learning approaches for landmark localization in medical images are often classification- \cite{ionasec2008dynamic, ibragimov2015computerized, dabbah_detection_2014, mahapatra_landmark_2012, lu_discriminative_2012, zheng2012automatic, urschler2018integrating, donner_global_2013, oktay2017stratified} or regression-based \cite{al2018automatic, stern_local_2016,han_robust_2014, gao_collaborative_2015, lindner2015fully, urschler2018integrating, donner_global_2013, oktay2017stratified}. Classification-based methods detect the presence of a landmark in image slices, patches, or voxels. Classification methods use a hard threshold: the landmark is either present or absent. Therefore, these methods usually rely on careful consideration of a final threshold value, which may be data and task specific. Regression-based methods circumvent the use of a hard threshold by outputting a continuous value\cite{zhou_discriminative_2014}. Regression-based methods predict the displacement or distance to the landmark from image slices, patches, or voxels. 
	
	Similar to many other automatic image analysis tasks, automatic landmark localization methods have become primarily deep learning-based \cite{litjens2017survey, zheng_3d_2015, o2018attaining}. Deep learning methods outperform conventional machine learning methods in a wide \REV{range} of applications\cite{litjens2017survey}. The advantage of deep learning is that it does not require handcrafting of features. 
	
	Several deep learning methods have been proposed for landmark localization that employ classification. Yang et al. \cite{yang_automated_2015} classified image slices with a convolutional neural network (CNN) and predicted a landmark location based on intersecting the classification outputs from all axial, coronal and sagittal image slices. Zheng et al. \cite{zheng_3d_2015} localized a landmark by classifying image voxels with multi-layer perceptrons, while Arik et al. \cite{arik2017fully} performed pixel classification with a CNN to localize landmarks. Xu et al. \cite{xu2017supervised} localized landmarks with a CNN that classified pixels based on their relative position (up, down, left or right) to the landmark of interest. Subsequently, landmarks were localized by using the obtained pixel-wise action steps.   
	
	Deep learning-based methods exploiting regression often predict heatmaps representing e.g. the distance between evaluated voxels and the landmark of interest \cite{wolterink2019coronary, payer2019integrating, o2018attaining, zhang2020headlocnet, torosdagli2018deep,  meyer2018pixel}. Landmarks are identified as local or global minima in these heatmaps. Voxel labels in heatmaps can be seen as pseudo-probabilities, indicating how close a voxel is located to a landmark. This makes heatmap regression comparable with voxel classification without using a hard threshold. Wolterink et al. \cite{wolterink2019coronary} employed a CNN containing dilated convolutions to predict heatmaps indicating landmark locations. Similar to Wolterink et al. \cite{wolterink2019coronary}, Payer et al. \cite{payer2019integrating} and O'Neil et al. \cite{o2018attaining} also proposed methods to predict heatmaps for automatic landmark localization. Payer et al.\cite{payer2019integrating} used a CNN that combined local appearance responses of a single landmark with the spatial configuration of that landmark to all other landmarks while O'Neil et al. \cite{o2018attaining} employed a CNN that analyzed low resolution images and subsequently used a second CNN for further refinement. Torosdagli et al. \cite{torosdagli2018deep} used a CNN to predict heatmaps representing the geodesic distance to a segmented organ containing landmarks and subsequently used a long short-term memory classification network to localize landmarks placed closely together. Unlike methods that performed a single task at a time, Meyer et al. \cite{meyer2018pixel} used a multi-task network to determine which landmark was closest to an analyzed pixel and subsequently predicted the normalized 2D distance towards that landmark. 
	
	Heatmap regression often requires combining a large number of predictions, for instance via a majority voting strategy, making it computationally expensive and often time-consuming. Therefore, a different approach was chosen by Zhang et al. \cite{zhang_detecting_2017}, who used a CNN to predict displacement vectors indicating the distance and direction from an analyzed voxel towards the landmark of interest. Subsequently, the CNN-architecture was expanded with additional layers to model correlations between analyzed input patches and output predicted landmark coordinates. Even though good results were obtained, predicting landmark coordinates directly from the image might be prohibited to large and complex CNN-architectures that model the complex non-linear mappings from input image to landmark location.
	
	Besides deep learning-based regression methods that directly predict landmark locations, regression has also been used to iteratively determine the landmark locations in an image\cite{aubert_automatic_2016, li2018fast, ghesu2019multi, alansary2019evaluating, al2019partial}. Aubert et al. \cite{aubert_automatic_2016} used a network to regress the displacement from an initial input patch, chosen with a statistical shape model, to the reference landmark. The landmark position was obtained by iteratively moving the input patch, using the predicted displacements, until convergence was reached and the landmark was localized. Li et al. \cite{li2018fast} localized landmarks in an iterative manner and employed a CNN that predicted the distance along each of the three coordinate axes from the center of 2.5D patches towards the landmark of interest while using classification to predict positive or negative movement along each coordinate axis. Ghesu et al. \cite{ghesu2019multi}, Alansary et al. \cite{alansary2019evaluating}, and Al et al. \cite{al2019partial} localized landmarks exploiting deep reinforcement learning to obtain the optimal search path from an initial starting location towards the landmark of interest. 

	\begin{figure}
		\centering	
		\includegraphics[width=\columnwidth]{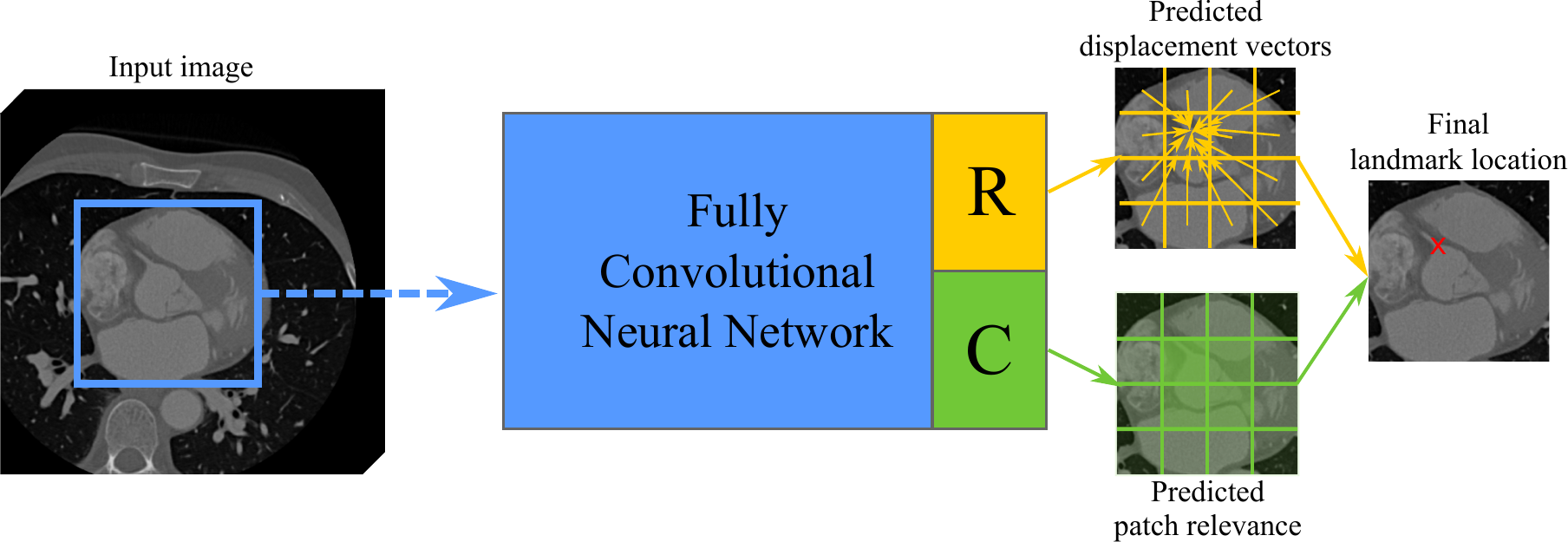}
		\caption{The fully convolutional neural network analyzes images in a patch-based manner, combining regression and classification. The landmark is localized by jointly predicting the displacement vector pointing from the center of each patch to the landmark with regression (R) and by predicting the presence of the landmark in each patch with classification (C). The final landmark location is obtained by computing a weighted average of the predicted displacement vectors, using the obtained posterior classification probabilities as weights during averaging.}
		\label{overview}	
	\end{figure}
	
	\begin{figure*}
	\centering	
		\includegraphics[width=\textwidth, trim=0cm 2.5cm 0cm 0cm]{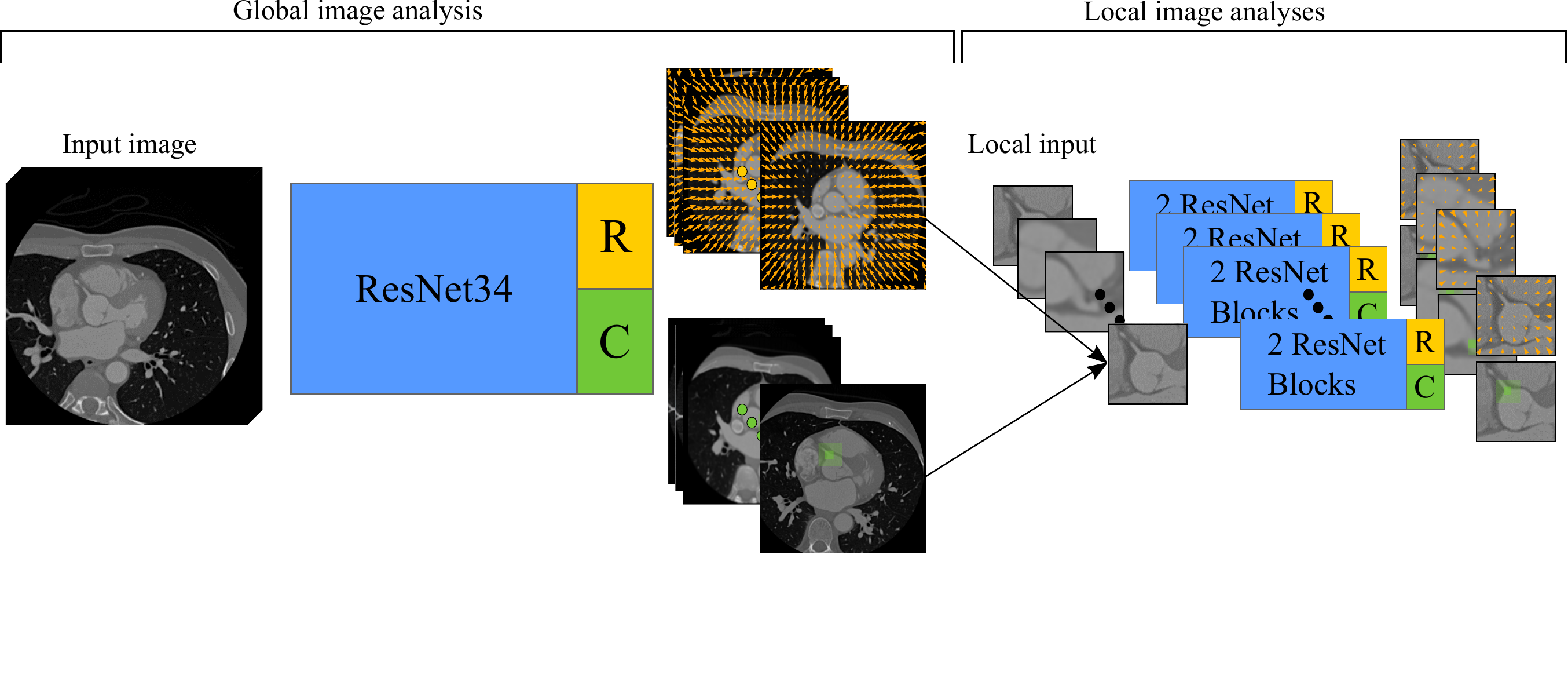}
		\caption{\REVV{Schematics of the proposed method for landmark localization. The method employs a global-to-local localization approach with ResNet-based CNNs. The first CNN provides global estimates of landmarks for the second specialized CNNs that predict final landmark locations. Since the CNNs are fully convolutional, they can handle input images of any size.}}
		\label{fig:network}	
	\end{figure*}
	
	\REVV{In this study, we propose a global-to-local localization approach, where an initial FCNN predicts the global locations of multiple landmarks simultaneously, and subsequently specialized FCNNs refine the final location of each landmark. Global multi-landmark localization and subsequent local single landmark localization are performed in a similar manner. While previous landmark localization methods used one approach, either classification \textit{or} regression, we propose a patch-based fully convolutional neural network (FCNN) that performs both classification \textit{and} regression (shown in Fig.~\ref{overview}). A patch-based approach provides a computationally efficient alternative to voxel-based approaches. However, since patch-based classification is inherently less precise than its voxel-wise counterpart, we mitigate this by jointly regressing the displacement vectors that point to the location of the landmark.} Conversely, using a regression-only localization approach might lead to sub-optimal localization results, because we postulate that displacement vectors predicted in image patches farther from the landmark of interest are less accurate than displacement vectors predicted in image patches closer to it. This can be mitigated by employing the posterior probabilities from the classification task as weights for weighted averaging of the displacement vectors. Combining regression and classification results in a landmark localization method that is both fast and highly accurate. We show that our method is generally applicable to a variety of landmark localization tasks:  8 landmarks in 3D coronary CT angiography (CCTA) scans, 2 landmarks in 3D olfactory MR scans, and 19 landmarks in 2D cephalometric X-rays. Additionally, we show that our method is able to localize single landmarks and multiple landmarks simultaneously.
	
	\section{Method}
	We propose an automatic landmark localization method that employs \REVV{a global-to-local estimation of landmark locations (Fig.~\ref{fig:network}). During global landmark localization, a fully convolutional neural network analyzes full input images in a patch-based manner and predicts the location of multiple landmarks. During subsequent local analysis, the location of each landmark is refined by an FCNN. The FCNNs employed for global and local analysis perform simultaneous regression and classification for a given input patch. In regression, the FCNNs predict displacement vectors from the center of any patch to landmarks of interest. The location of each landmark might be obtained by computing the average of the landmark locations to which predicted displacement vectors point, but presumably not all patches are equally important for accurate localization; i.e. image patches closer to the target landmark are likely more relevant for accurate landmark localization than patches farther from the target landmark. Therefore, simultaneously to regression, classification is performed to determine the importance of each patch in the landmark localization. Classification determines the presence of the target landmark in an image patch and the obtained posterior classification probabilities are used for weighted averaging of all predicted displacement vectors.}
	
	\REVV{The global FCNN is based on ResNet34~\cite{he2016deep} and it consists of one convolutional layer with 16 ($7\times7\times7$) kernels and a stride of 2, which is followed by 4 ResNet-blocks. One ResNet-block contains 3, 4, or 6 convolutional layer pairs, where each convolutional layer pair consist of two convolutional layers with 32, 64, 128, or 256 ($3\times3\times3$) kernels. In contrast to the original ResNet34\cite{he2016deep}, which contains a strided convolutional layer as first layer in every ResNet-block, our network contains a pooling layer before the first and second ResNet-block, which is an average pooling layer with a size and stride of $2\times2\times2$ voxels. After the four ResNet-blocks, the network has two output heads: one for regression of displacement vectors, and another for classification of landmark presence. Both output heads are similar in design. Each head has two 256-node dense layers and an output-layer, implemented as 1$\times$1$\times$1 convolutions \cite{long_fully_2015}. The classification head outputs scalars, one for each landmark, forced between 0 and 1 by a sigmoid function. The regression head outputs displacement vectors for each landmark.}

    \REVV{The specialized FCNNs for local landmark prediction are of similar but smaller design. Each network consists of a ResNet-block, followed by average pooling, a second ResNet-block, and the parallel regression and classification heads. The first ResNet-block consists of two convolutional layers of 32 ($3\times3\times3$) kernels. Average pooling is done with a size and stride of $2\times2\times2$ voxels. The second ResNet-block consists of two convolutional layers with 64 ($3\times3\times3$) kernels. Similarly to the global FCNN, the two ResNet-blocks are followed by two output heads: one for the classification task, and another for the regression task. All layers use 64 kernels.}
	
    Each convolutional layer in the FCNNs applies zero-padding, and after each convolutional layer, batch normalization \cite{ioffe_batch_2015} is applied. To allow application to images of arbitrary size, 3D feature maps are not flattened but dense layers are implemented as convolutions with a size of 1$\times$1$\times$1 voxel\cite{long_fully_2015}. Throughout a network, \REVV{rectified linear units (ReLUs)} are used for activation, except for the regression and classification output layers. For regression, a linear activation function is used, and for classification, a sigmoid activation is used to obtain posterior probabilities between~0~and~1.
    
    The loss function that was optimized during training consisted of two parts: the mean absolute error between the regression output and reference displacements, and the binary cross-entropy between the classification output and reference labels. To ensure that input patches located far from the landmark have less influence on updates of network parameters compared to those located close to the landmark, the mean absolute error is calculated on log-transformed displacement vectors. As optimization algorithm, Adam with a learning rate of 0.001 \cite{kingma_adam:_2014} was used.
	
	Since a network is fully convolutional it can analyze input images of varying size. Depending on its input image the network outputs a varying number of displacement vectors and posterior probabilities during global landmark localization. Due to the network's \REVV{average pooling layers and the first convolutional layer with a stride of two voxels}, its outputs are distributed on a grid, where the grid spacing is defined by \REVV{the sum of the number of pooling layers and strided convolutional layers}. With $n$ representing \REVV{the sum of the number of pooling layers and strided convolutional layers for the global or local localization step}, this leads to a down-sampling rate of $1/2^n$ and therefore, a patch size of $2^n$ voxels. Hence, for a given network, a grid with a grid spacing of $2^n$ voxels is used to sample patches from an input image. 

	\section{Data}
	\begin{figure}[!]
		\centering
		\begin{tabular}{@{}c@{}}
			\includegraphics[width=0.485\textwidth]{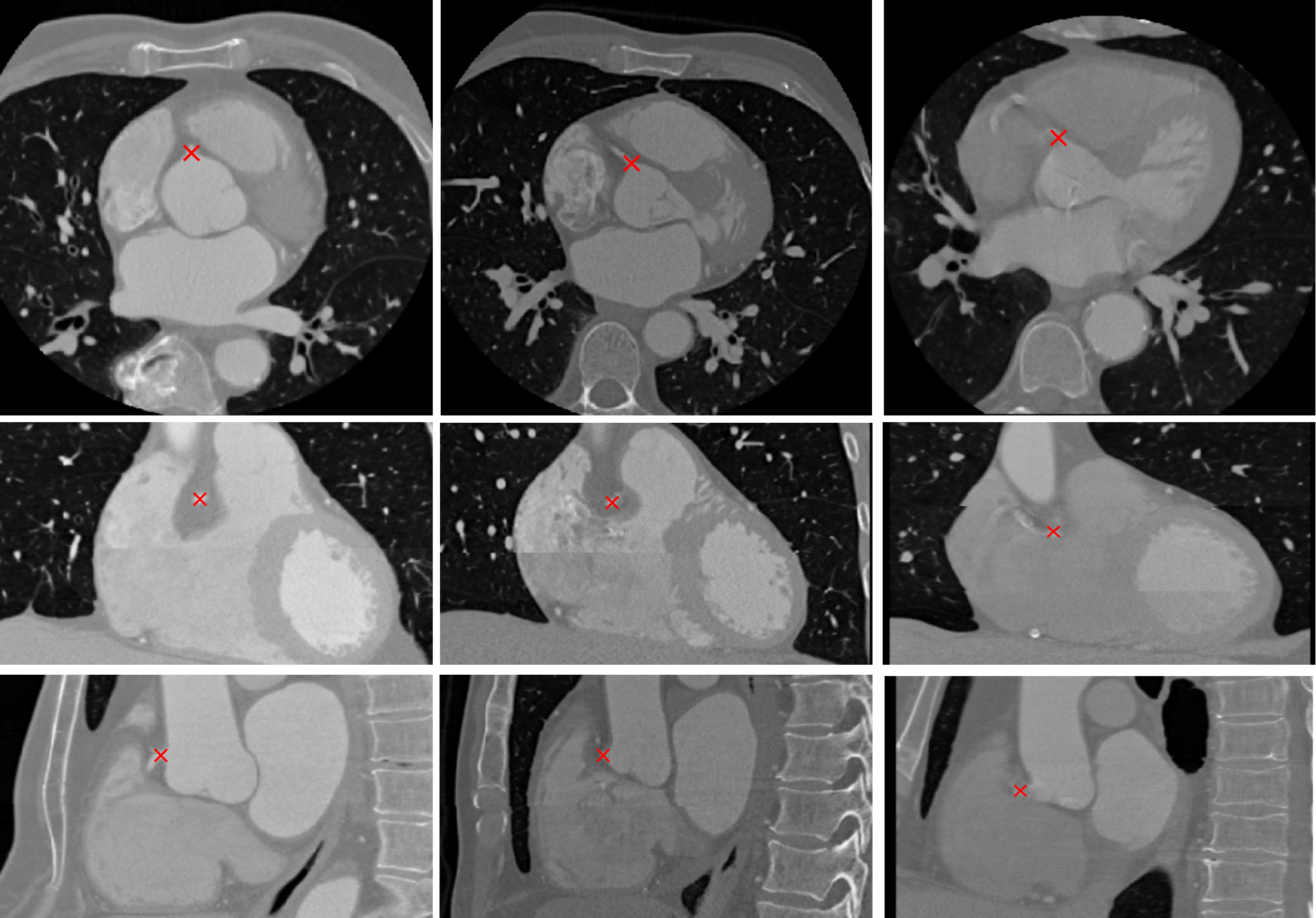}\\
			\small (a) CCTA 
		\end{tabular}
		\vspace{\floatsep}
		
		\begin{tabular}{@{}c@{}}
			\includegraphics[width=0.485\textwidth]{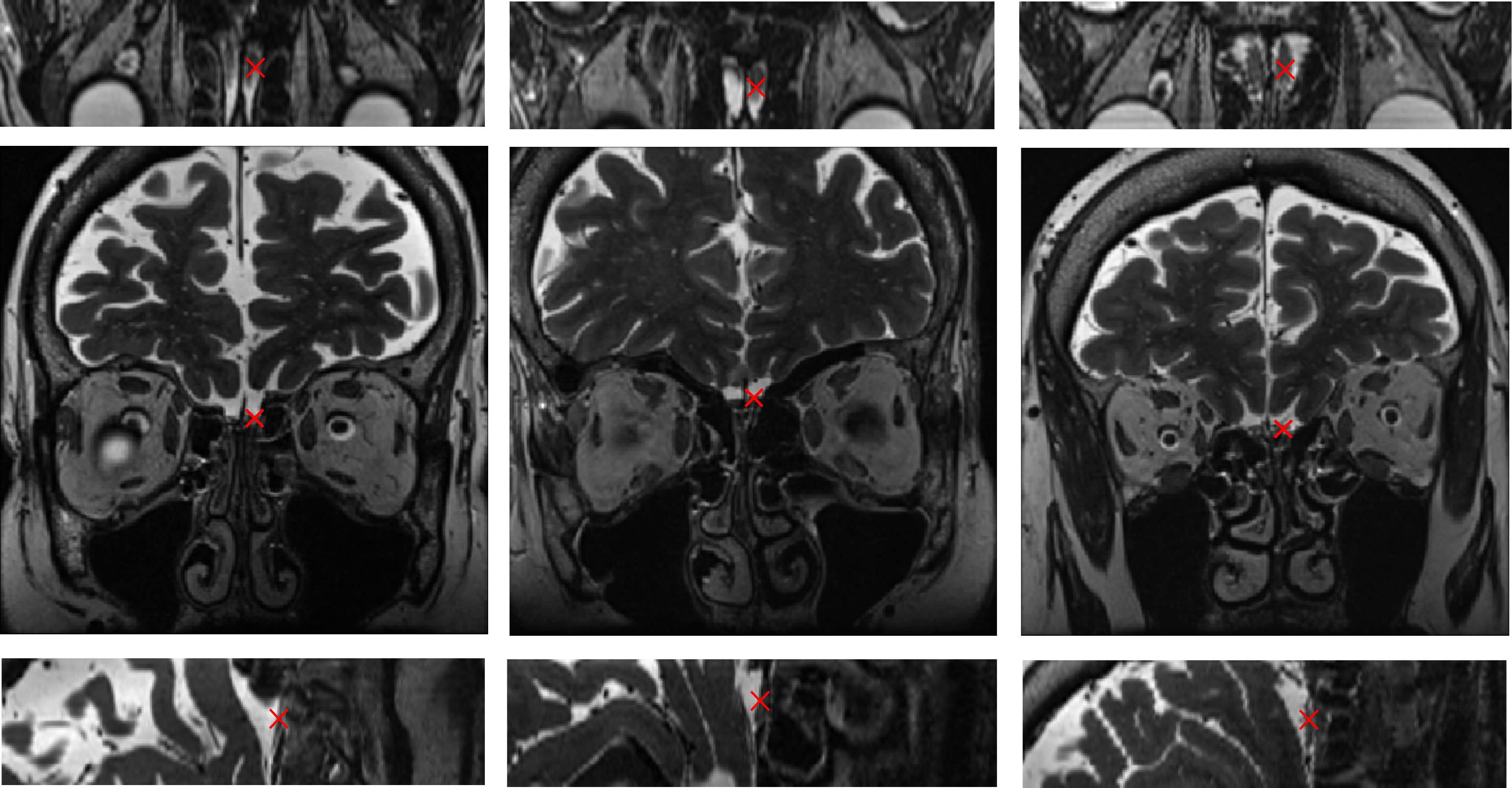}\\
			\small (b) Olfactory MR
		\end{tabular}	
		\vspace{\floatsep}
		
		\begin{tabular}{@{}c@{}}
			\includegraphics[width=0.485\textwidth]{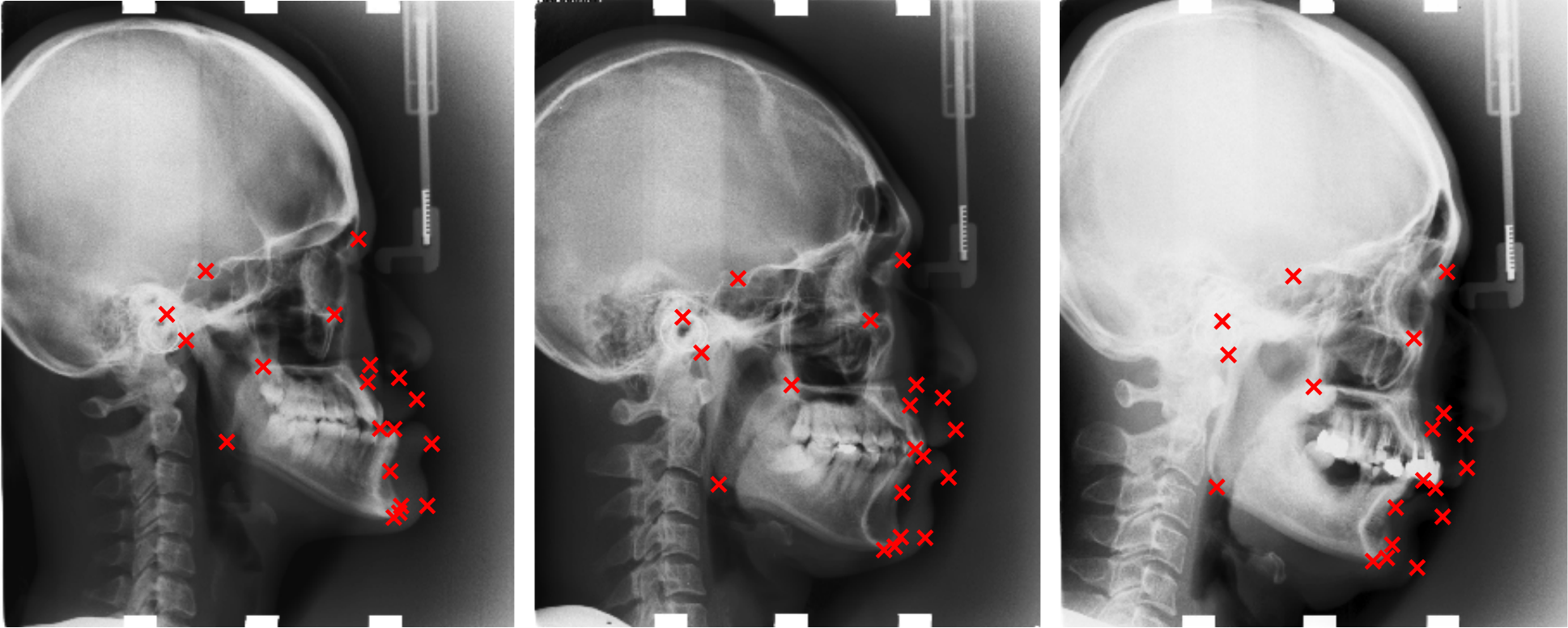}\\
			\small (c) Cephalometric X-rays
		\end{tabular}
		\caption{Example images with reference annotations. (a) Axial, coronal, and sagittal slices (rows) from cardiac CT angiography (CCTA) scans of three different patients (columns), in which the right ostium is indicated by a red cross. (b) Axial, coronal, and sagittal slices (rows) from olfactory MR scans of three different patients (columns), in which the center of the right olfactory bulb is indicated by a red cross. (c) Cephalometric X-ray of three different patients (columns), in which 19 different landmarks are indicated by a red cross.}	
		\label{fig:GT_annotations}
	\end{figure}
	We evaluated the method on three different datasets containing 3D CCTA scans, 3D olfactory MR scans, and 2D cephalometric X-rays (Fig.~\ref{fig:GT_annotations}). The choice of these datasets was based on the diversity in image acquisition modality (CT, MR, and X-ray), image dimensionality (2D and 3D), and anatomical coverage (cardiac, brain, and head).
	
	\subsection{Coronary CT Angiography}
	The dataset consisted of 672 CCTA scans, which were acquired in the University Medical Center Utrecht (Utrecht, The Netherlands) as part of regular patient care. The need for informed consent was waived by the Institutional Medical Ethical Review Board. ECG-triggered scans were acquired with a 256-detector row scanner (Philips Brilliance iCT, Philips Medical, Best, The Netherlands). Tube voltage ranged from 80~to~140\,kVp while tube current ranged from 210~to~300\,mAs. Intravenous contrast was administered before acquisition. All acquired scans had a slice thickness of 0.9\,mm with 0.45\,mm spacing. In-plane resolution ranged between 0.29 and 0.49\,mm. In all scans, an expert manually annotated eight clinically relevant cardiac landmarks: the aortic valve commissures between the non-coronary and right (NCRC), the non-coronary and left (NCLC), and the left and right aortic valve leaflets (LRC), the hinge points (most caudal attachments) of the left (LH), non-coronary (NCH), and right (RH) aortic valve leaflets, and the right (RO) and left coronary ostium (LO). These landmarks can be used to perform clinical measurements in patients undergoing transcatheter aortic valve implantation (TAVI) \cite{kasel2013standardized, elattar2016automatic, zheng2012automatic, wachter2010patient}. \REVV{An expert observer created manual annotations for the full dataset using the protocol by Kasel et al.~\cite{kasel2013standardized}. To determine inter-observer variability, a second observer annotated 100 randomly selected scans from the test-set. This same set was used to determine the intra-observer variability: after one month, the first observer repeated annotations in the 100 scans. Variability was defined as the Euclidean distance between the landmark annotations.}
	
	\subsection{Olfactory MR}
	The dataset contained 61 olfactory MR scans, which were acquired as part of clinical routine in Hospital Gelderse Vallei (Ede, The Netherlands). The local ethics committee approved the study where informed consent was obtained from all subjects. Scans were acquired with a 3T Magnetom Verio MRI scanner (Siemens, Erlangen, Germany). To visualize the olfactory bulbs, a coronal T2-weighted fast spin-echo sequence was performed (echo time: 153 ms, repetition time: 4630 ms, field of view: 120$\times$120\,mm). Each scan contained 28 coronal slices reconstructed to an isotropic in-plane pixel size of 0.47\,mm and a slice thickness of 1\,mm with no inter-slice gap. In each scan, an expert manually delineated the right and left olfactory bulb. The center of each manual delineation was taken as ground truth landmark location. 
	
	\subsection{Cephalometric X-rays}
	The dataset consisted of 400 publicly available cephalometric X-rays (lateral cephalograms) from the \textit{ISBI 2015 Grand Challenge in Automatic Detection and Analysis for Diagnosis in Cephalometric X-ray Images} \cite{wang2016benchmark}. X-rays were acquired with a Soredex CRANEX\textregistered \ Excel Ceph machine (Tuusula, Finland) and Soredex SorCom software (3.1.5, version 2.0), and were obtained in TIFF format with a resolution of $1935\times2400$ pixels and an isotropic pixel size of 0.1\,mm. In all X-rays, two experienced medical doctors both manually annotated 19 clinically relevant landmarks, which can be used for diagnosis and treatment planning in orthodontic patients \cite{wang2016benchmark, torosdagli2018deep}. Following the challenge protocol, the average of the annotations provided by both experts was used as ground truth landmark location \cite{wang2016benchmark}. The landmarks were: the sella (L1), nasion (L2) orbitale (L3), porion (L4), subspinale (L5), supramentale (L6), pogonion (L7), menton (L8), gnathion (L9), gonion (L10), lower incisal incision (L11) upper incisal incision (L12), upper lip (L13), lower lip (L14), subnasale (L15), soft tissue pogonion (L16), posterior nasal spine (L17), anterior nasal spine (L18), and articulate (L19). The intra-observer and interobserver variability were determined within the challenge following Lindner et al.\cite{lindner2015fully}. 
	
	\section{Evaluation}
	Evaluation was performed by computing the median Euclidean distance and interquartile range (IQR) between manually defined reference and automatically predicted landmark locations. 
	
	In addition, following the \textit{ISBI 2015 Grand Challenge in Automatic Detection and Analysis for Diagnosis in Cephalometric X-ray Images} \cite{wang2016benchmark}, success detection rates (SDRs) were calculated. The detection of a landmark is considered successful when the Euclidean distance between the automatically localized landmark and its reference location is smaller than a predefined distance threshold. In our analysis we used 10 distance thresholds ranging from 0.5\,mm to 5\,mm. The maximal distance threshold was defined when 95\% of the landmarks were successfully detected. \REVV{Intra- and second observer SDRs were determined in a similar way by using the two annotated sets. When two annotations of a landmark were within the distance of the set threshold, the annotation was considered successful.}
	
	\section{Experiments and Results}
	The method was implemented in Python using \REVV{PyTorch \cite{NEURIPS2019_9015} on an NVIDIA 2080 Ti with 11 GB of memory.}

	\subsection{Experiments}
	All datasets were first randomly divided into a training set, a validation set, and a hold-out test set. Training sets and validation sets were used to develop the method, while test sets were used for final evaluation. Note that the test sets were not used during method development in any way. 
	
	\subsubsection{Coronary CT Angiography}
	\label{sec:ccta}
	
	The available set of 672 scans was randomly divided into 412 training, 60 validation and 200 test scans. For computational purposes, scans were resampled to an isotropic voxel size of 1.5\,mm$^3$. 
	
	\REVV{The global FCNN was trained for 300,000 iterations, using mini-batches of 4 randomly sampled sub-images of $72\times72\times72$ voxels. The FCNN was evaluated during training on the entire validation set every 10,000 iterations. The best performing model was used for subsequent analysis. A local FCNN was trained using similar settings. However, the size of the sub-images was chosen based on the distance errors obtained during localization of landmarks with the multi-landmark network in the validation set and was therefore set to $16\times16\times16$ voxels. Moreover, sub-images were randomly sampled such that they always contained the landmark of interest.}

\begin{table*}
 	\setlength{\tabcolsep}{5pt}
		\renewcommand{\arraystretch}{1.3}
		\centering
		\caption{
		\REV{\REVV{Median (IQR) Euclidean distance errors (mm) between computed and reference landmark locations in CCTA scans obtained with networks used for multi-landmark localization. Different training settings are evaluated: regression of displacement vectors with (R\textsubscript{log}) and without (R) log-transformation, employing the classification layer (C) or not, and performing only global localization (Global) or global-to-local localization (Global-to-local) of landmarks. The proposed method combined global-to-local landmark localization and employed regression of log-transformed displacement vectors and the classification output layer (Proposed ML). Additionally, distance errors obtained with the method adjusted for single landmark localization are listed as well (Proposed SL). The distance errors obtained by the intra-observer (Intra-observer) and the second observer (Second observer), computed as the distance between two annotations made by the same observer, and the distance between annotations made by two different observers, respectively, on a subset of the test set are also listed. Results are listed per landmark (NCRC, NCLC, LRC, LH, NCH, RH, RO, and LO) and for all landmarks together (All), with the smallest distance error shown in \textbf{bold}}.}}
		\begin{tabular}{l|cccccccc|c}
			& \multicolumn{3}{c}{Aortic valve commissures} & \multicolumn{3}{c}{Aortic valve hinges} & \multicolumn{2}{c|}{Coronary ostia} & \\ 
			& NCRC        & NCLC        & LRC          & LH          & NCH         & RH          & RO          & LO          & All       \\\hline
			
			Intra-observer & 2.68 (2.25) & 1.93 (1.72) & 1.96 (2.07)\ssymbol{2} & 2.04 (1.39)* & 2.54 (2.20) & 2.56 (2.28)\ssymbol{3} &  1.43 (1.05) & 1.88 (1.40)* & 2.06 (1.84)\ssymbol{3}\\
			Second observer & 3.00 (1.23)\ssymbol{3} & 3.46 (1.45)\ssymbol{3} & 2.96 (1.13)\ssymbol{3} & 1.73 (1.32)* & 1.96 (1.19)\ssymbol{3} & 1.80 (1.62) & 1.78 (1.54)\ssymbol{2} & 2.31 (1.56)\ssymbol{3} & 2.50 (1.68)\ssymbol{3} \\\hline
            \textit{Global} \\
		    R & \REVV{3.20 (2.05)\ssymbol{3}} & \REVV{2.95 (1.95)\ssymbol{2}} & \REVV{3.08 (2.47)\ssymbol{3}} & \REVV{3.14 (2.10)\ssymbol{3}} & \REVV{3.51 (2.04)\ssymbol{3}} & \REVV{3.55 (2.14)\ssymbol{3}} & \REVV{4.36 (2.92)\ssymbol{3}} & \REVV{3.92 (2.71)\ssymbol{3}} &\REVV{ 3.44 (2.36)\ssymbol{3}}  \\
		    R\textsubscript{log} & \REVV{3.12 (2.06)\ssymbol{3}} & \REVV{3.12 (2.02)\ssymbol{3}} & \REVV{3.04 (2.09)\ssymbol{3}} & \REVV{2.49 (1.84)} & \REVV{3.36 (2.18)\ssymbol{3}} & \REVV{3.18 (1.71)\ssymbol{3}} & \REVV{4.08 (2.87)\ssymbol{3}} & \REVV{4.05 (2.66)\ssymbol{3}} & \REVV{3.23 (2.30)\ssymbol{3}}\\
		    C & \REVV{5.64 (4.13)\ssymbol{3}} & \REVV{4.77 (2.10)\ssymbol{3}} & \REVV{4.62 (2.37)\ssymbol{3}} & \REVV{4.43 (2.23)\ssymbol{3}} & \REVV{4.69 (3.01)\ssymbol{3}} & \REVV{4.34 (2.73)\ssymbol{3}} & \REVV{5.07 (2.67)\ssymbol{3}} & \REVV{4.64 (2.61)\ssymbol{3}} & \REVV{4.76 (2.72)\ssymbol{3}}  \\
		    R + C & \REVV{2.93 (1.96)} & \REVV{2.60 (1.53)} & \REVV{2.62 (2.54)} & \REVV{2.73 (1.54)\ssymbol{2}} & \REVV{3.31 (1.92)\ssymbol{2}} & \REVV{3.14 (1.66)\ssymbol{2}} & \REVV{4.23 (2.94)\ssymbol{3}} & \REVV{3.72 (2.78)\ssymbol{2}} & \REVV{3.09 (2.11)\ssymbol{3}}   \\
		    R\textsubscript{log} + C & \REVV{ 2.72 (1.71)} & \REVV{2.59 (1.84)} & \REVV{2.60 (1.98)} & \REVV{2.51 (1.74)} & \REVV{3.08 (1.79)} & \REVV{2.77 (1.70)} & \REVV{2.90 (1.93)} & \REVV{3.31 (2.23)} & \REVV{2.81 (1.88)} \\\hline
            \textit{Global-to-local} \\
		    Proposed SL & \REVV{1.99 (1.82)} & \REVV{1.94 (1.32)} & \REVV{\textbf{1.69 (1.73)}} & \REVV{\textbf{2.29 (1.46)}} & \REVV{2.68 (1.75)\ssymbol{3}} & \REVV{3.09 (1.82)\ssymbol{3}} & \REVV{1.48 (0.98)} & \REVV{1.55 (1.02)} & \REVV{2.03 (1.79)\ssymbol{3}} \\ 
		    Proposed ML & 
		    \REVV{\textbf{1.85 (1.96)}} & \REVV{\textbf{1.80 (1.59) }}& \REVV{1.76 (1.67)} & \REVV{2.40 (1.58)} & \REVV{\textbf{2.48 (1.72)}} & \REVV{\textbf{2.23 (1.42)}} & \REVV{\textbf{1.45 (1.20)}} & \REVV{\textbf{1.55 (0.97)}} &\REVV{\textbf{ 1.87 (1.67)}} \\ 
			
		\end{tabular}
		 \begin{tablenotes}
         \small
         \item Significance outcome by the Wilcoxon Signed Rank test compared to R\textsubscript{log} + C for the upper part of the table, and the proposed method for multi-landmark localization (Proposed ML) and the Intra-observer, Second observer, and proposed method for single landmark localization (Proposed SL) is indicated with * for p<0.05, $\dagger$ for p<0.01, and $\ddagger$ for p<0.001.
         \end{tablenotes}
		\label{tab:ccta_results}
 	\end{table*}

\begin{figure*}
		\centering	
		\includegraphics[]{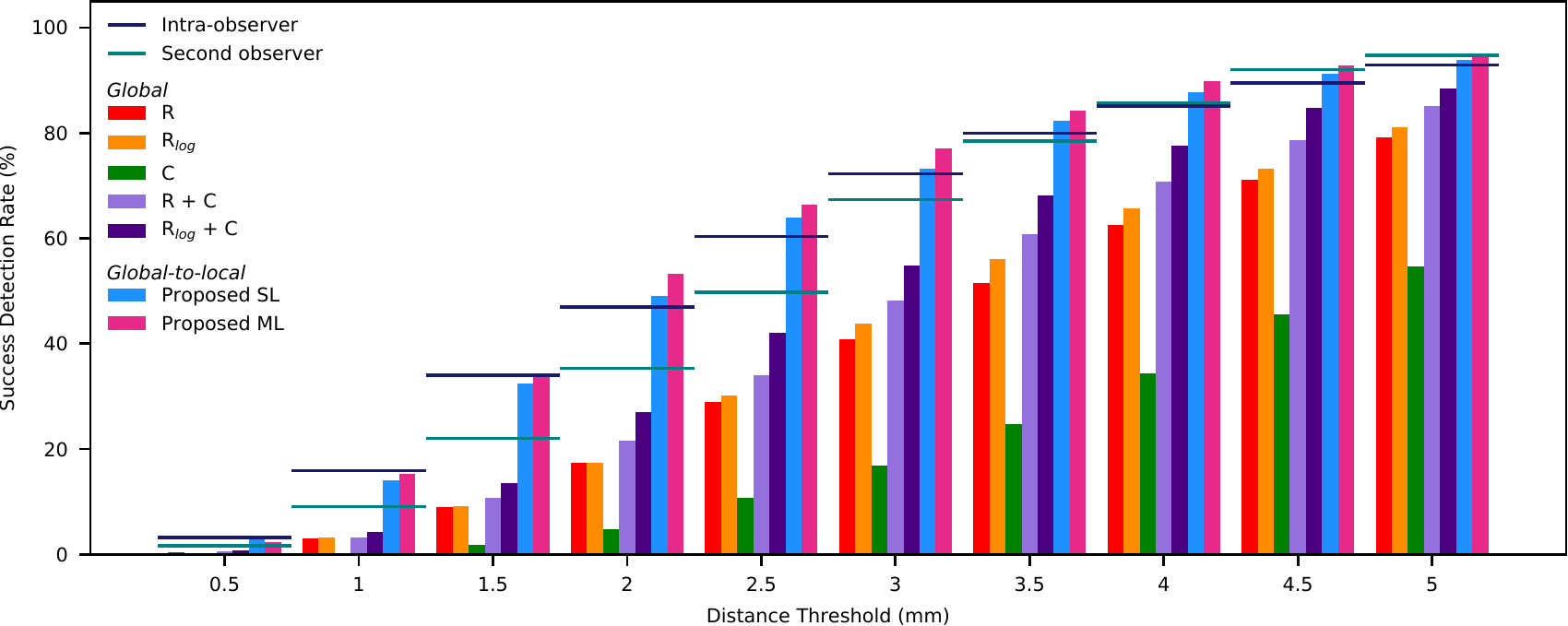}
		\caption{\REVV{Success detection rates (SDRs) for landmark localization in CCTA scans. The results for the ablation study on the performance of global multi-landmark localization (Global) are shown for: FCNNs trained for one task, i.e. regression of displacement vectors (R), regression of a log-transformed displacement vectors (R\textsubscript{log}), and classification of patches (C); FCNNs trained for joint regression and classification (R + C) and joint regression of log-transform displacement and classification (R\textsubscript{log} + C). Furthermore, the results for our proposed global-to-local FCNNs (Global-to-local) trained for single landmark localization (Proposed SL) and multi-landmark localization (Proposed ML) are also shown. Additionally, intra-observer and second observer SDRs are indicated by horizontal lines. Results are shown as \% over all landmarks.}}
		\label{fig:CCTA_sdr_method}	
	\end{figure*}

	Table \ref{tab:ccta_results} lists the obtained median Euclidean distance errors (last row: Proposed ML) and those obtained by the intra-observer and second observer (first two rows) per landmark and for all landmarks together. Median distance errors obtained with the proposed method range from 1.45 to 2.48 mm for different landmarks, which corresponds to an error between 1.03 and 1.65 voxels. \REV{This is close to \REVV{distance errors obtained by the intra-observer} which ranged from 1.43 to 2.68 mm, and \REVV{the distance errors obtained by the second observer} which ranged from 1.73 to 3.46 mm.} Distance errors obtained for automatic localization of the coronary ostia were the smallest, with 1.45 mm for the RO and 1.55 mm for the LO. On average, the processing time per scan was 0.06~$\pm$~0.05 seconds. \REVV{For six out of eight landmarks, distance errors obtained with the proposed method were lower than distance errors obtained by the intra-observer annotation. For three of these six landmarks differences were statistically significant. For five out of eight landmarks, distance errors obtained with the proposed method were lower than distance errors obtained by the second observer. For all landmarks but the RH, the differences were statistically significant}. To provide further insight in the performance, Fig. \ref{fig:CCTA_sdr_method} shows the SDRs obtained with the proposed method, while the intra-observer and second observer SDRs are shown as horizontal lines. Overall, the SDRs obtained with the method are \REVV{similar or better than intra-observer and second-observer SDRs}. 
	
\begin{figure}
		\centering
		\includegraphics[width=\columnwidth]{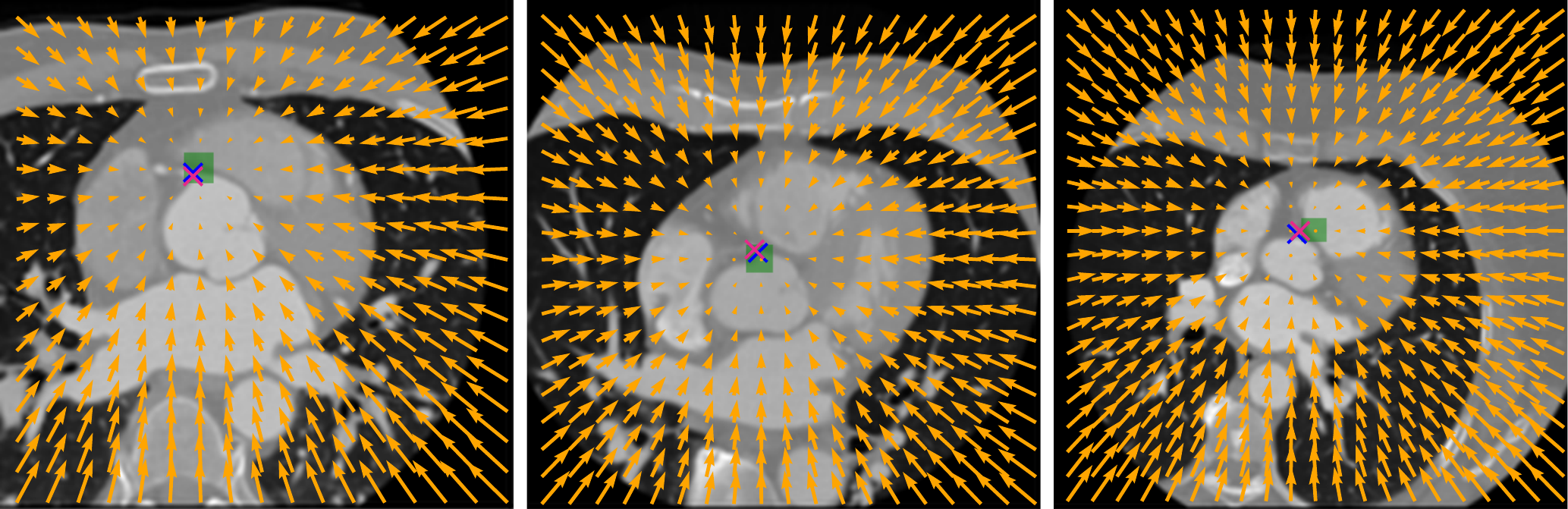}\\
		\vspace{\floatsep}
		\includegraphics[width=\columnwidth]{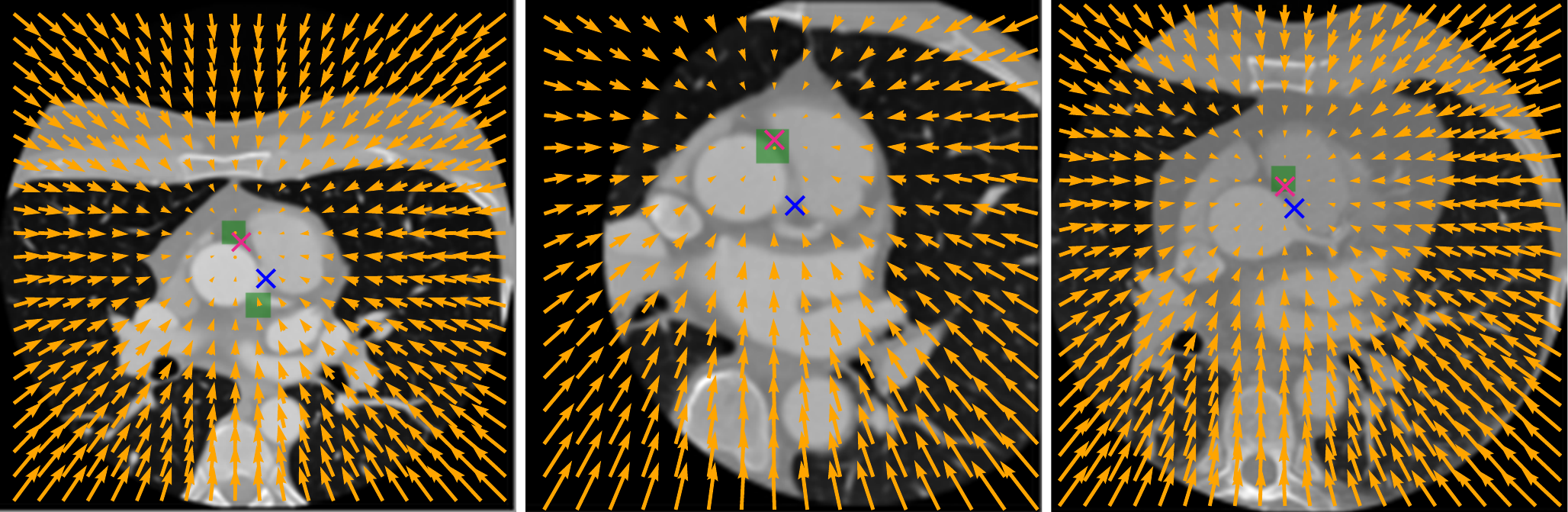}\\
		\caption{Vector fields (orange) visualizing the predicted displacement vectors in the axial plane in six different CCTA scans from the test set where localization of the right coronary ostium was performed. For visualization purposes, 3D predicted displacement vectors are shown as 2D vector fields and the magnitudes of the vectors are rescaled. The green squares indicate posterior probabilities larger than 0.5, obtained by the classification task of the network. Reference and computed landmark locations are indicated with a blue and pink cross, respectively. The top row depicts scans in which localization errors were below 2.5 mm, while the bottom row depicts scans in which localization errors were above 5.0 mm. Images in the bottom row all contain two coronary ostia which are both located on the left side in close proximity of each other.}
		\label{fig:vectorsCCTA}	
	\end{figure}

	Fig. \ref{fig:vectorsCCTA} shows vector fields visualizing the predicted displacement vectors for localization of the RO landmark in the axial viewing plane in six CCTA scans from the test set: three scans in which the localization error was below 2.5 mm (top row) and three scans in which the localization error was above 5.0 mm (bottom row). Larger errors were made in scans in which anatomical deviation was present, such as both coronary ostia being located on the left side in close proximity to each other (Fig. \ref{fig:vectorsCCTA} bottom row). This anatomical deviation occurred in only 0.2\% of the scans in the training set but in 2\% of the scans in the test set, which might explain the error.
	
	\subsubsection{Olfactory MR}
	The available set of 61 olfactory MR scans was randomly divided into 36 training, 5 validation and 20 test scans. Scans were resampled to an isotropic voxel size of 0.47\,mm$^3$. \REVV{Training settings were similar to those used in the CCTA experiment, described in Section~\ref{sec:ccta}.} However, because scans contained only 60 coronal slices after resizing, scans were zero-padded in the z-direction. The size of the olfactory bulbs ranged between 1.4 and 5.2\,mm in-plane, and between 5.0 and 13.0\,mm in the z-direction. 
	
\begin{figure}
		\centering	
		\includegraphics[width=\columnwidth]{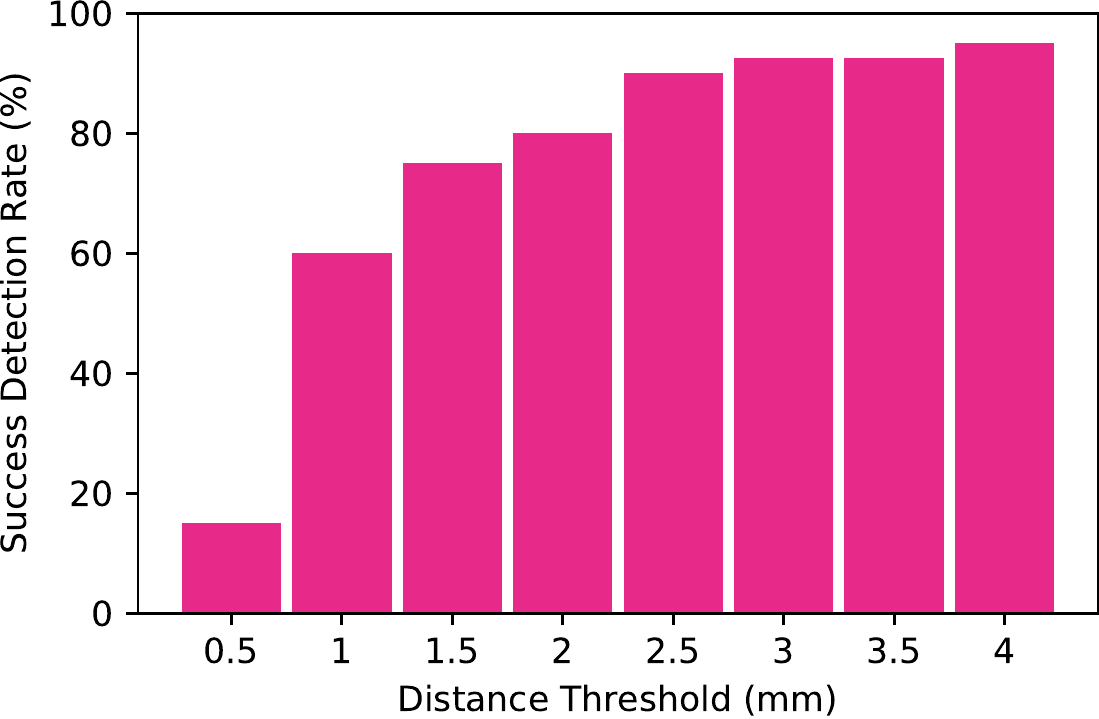}
		\caption{\REVV{Success detection rates (SDRs) for olfactory bulb localization in MR scans, obtained with the proposed method.  Results are shown as \% over all landmarks present in the test set and are given for eight distance thresholds ranging from 0.5\,mm to 4\,mm.}}
		\label{fig:bulb_sdr}	
	\end{figure}

	The median (IQR) Euclidean distance error between computed landmark locations and reference locations was \REVV{0.87 (1.36)} mm and \REVV{0.90 (0.58)} mm for the right and left bulb, respectively, and \REVV{0.90 (0.85)} mm when taking both landmarks into account. Fig.~\ref{fig:bulb_sdr} shows the SDRs obtained with the proposed method. When a distance threshold of 4 mm was used, 95.0\% of all landmarks present in the test set were successfully detected. The processing time per scan was on average \REVV{0.07~$\pm$~0.01} seconds.
	
	\subsubsection{Cephalometric X-rays}
	The 150 cephalometric X-rays from the training set of the \textit{ISBI 2015 Grand Challenge in automatic Detection and Analysis for Diagnosis in Cephalometric X-ray Images} \cite{wang2016benchmark} were used for training (140 X-rays) and validation (10 X-rays). The challenge provides two separate test sets for evaluation: one set containing 150 images (Test1), and one set containing 100 images (Test2). To mitigate varying image contrast, histogram equalization was performed on the X-ray images before analysis. Since the X-ray is a 2D image, we used a 2D version of the network in Fig.~\ref{fig:network}. Furthermore, because cephalometric X-rays are large ($1935\times2400$ pixels), we \REVV{also added an average pooling layer before the third and fourth ResNet-block}. Adding \REVV{average} pooling layers allowed us to enlarge the receptive field, while keeping a low computational complexity. \REVV{The network for global localization was again trained for 300,000 iterations, using mini-batches containing 4 sub-images of $592\times592$ pixels. For local analysis, mini-batches containing 4 sub-images of $16\times16$ pixels were used during training}. 
	
	Table \ref{tab:ceph_table} lists the median Euclidean distance errors obtained with the proposed method. Errors range from \REVV{0.46 to 2.12 mm} for different landmarks in Test1 and from \REVV{0.42 to 4.32 mm} for different landmarks in Test2. For both test sets, the best results were obtained for the localization of L12, which is the upper incisal incision. As reported by Lindner et al. \cite{lindner2015fully}, the mean intra-observer variability for the first and second observer were 1.73~$\pm$~1.35\,mm and 0.90~$\pm$~0.89\,mm, respectively, while the mean inter-observer variability was 1.38~$\pm$~1.55\,mm. When computing the mean distance error obtained on all landmarks present in both test sets, we obtain a distance error of \REVV{1.35~$\pm$~1.19\,mm, which is lower than the intra-observer variability of the first observer and the inter-observer variability}. As defined by the challenge protocol \cite{wang2016benchmark}, we evaluated our method computing the SDRs using four distance thresholds (2\,mm, 2.5\,mm, 3\,mm, and 4\,mm). These results are shown in Fig. \ref{fig:ceph_comparison}. On average, the processing time per scan was 0.05~$\pm$~0.009 seconds.
    
	\afterpage{\begin{table*}
		\renewcommand{\arraystretch}{1.3}
		\centering
		\caption{Median (IQR) Euclidean distance errors (mm) between the computed landmark locations and the reference locations, obtained with the proposed method. Results are listed separately for the two different test sets: Test1 and Test2. Furthermore, results are listed per landmark (L1-L19) and for all landmarks together (All).}
		\begin{adjustbox}{}
			\begin{tabular}{lcccccccccc}
				& L1          & L2          & L3          & L4          & L5          & L6           & L7          & L8          & L9          & L10         \\\hline
				Test1 & \REVV{0.51 (0.41)} & \REVV{1.00 (1.25)} & \REVV{1.01 (1.09)} & \REVV{1.62 (1.71)} & \REVV{1.64 (1.64)} & \REVV{0.94 (0.98)} & \REVV{0.70 (0.85)} & \REVV{0.63 (0.70)} & \REVV{0.76 (0.85)} & \REVV{2.12 (1.83)} \\
				
				Test2 & \REVV{0.52 (0.34)} & \REVV{0.57 (1.00)} & \REVV{2.31 (1.40)} & \REVV{1.19 (1.49)} & \REVV{1.11 (1.06)} & \REVV{2.62 (1.73)} & \REVV{0.58 (0.72)} & \REVV{0.50 (0.47)} & \REVV{0.52 (0.55)} & \REVV{1.68 (1.61)}\\
		
				& L11         & L12         & L13         & L14         & L15         & L16          & L17         & L18         & L19         & All       \\\cline{1-11}
				Test1 & \REVV{0.80 (1.26)} & \REVV{0.46 (0.89)} & \REVV{1.13 (0.83)} & \REVV{0.84 (0.58)} & \REVV{0.90 (0.88)} & \REVV{1.23 (1.14)} & \REVV{0.64 (0.64)} & \REVV{0.94 (1.21)} & \REVV{1.50 (1.78)} & \REVV{0.95 (1.15)} \\
				Test2 & \REVV{0.63 (0.87)} & \REVV{0.42 (0.67)} & \REVV{2.32 (0.87)} & \REVV{1.87 (1.09)} & \REVV{0.94 (0.64)} & \REVV{4.32 (1.47)} & \REVV{0.88 (0.81)} & \REVV{1.13 (1.19)} & \REVV{1.06 (1.35)} & \REVV{1.07 (1.60)}
			\end{tabular}
		\end{adjustbox}
		\label{tab:ceph_table}
	\end{table*}}
	
	\afterpage{\begin{figure*}
		\centering
		\begin{tabular}{@{}c@{}}
			\includegraphics[width=0.32\textwidth]{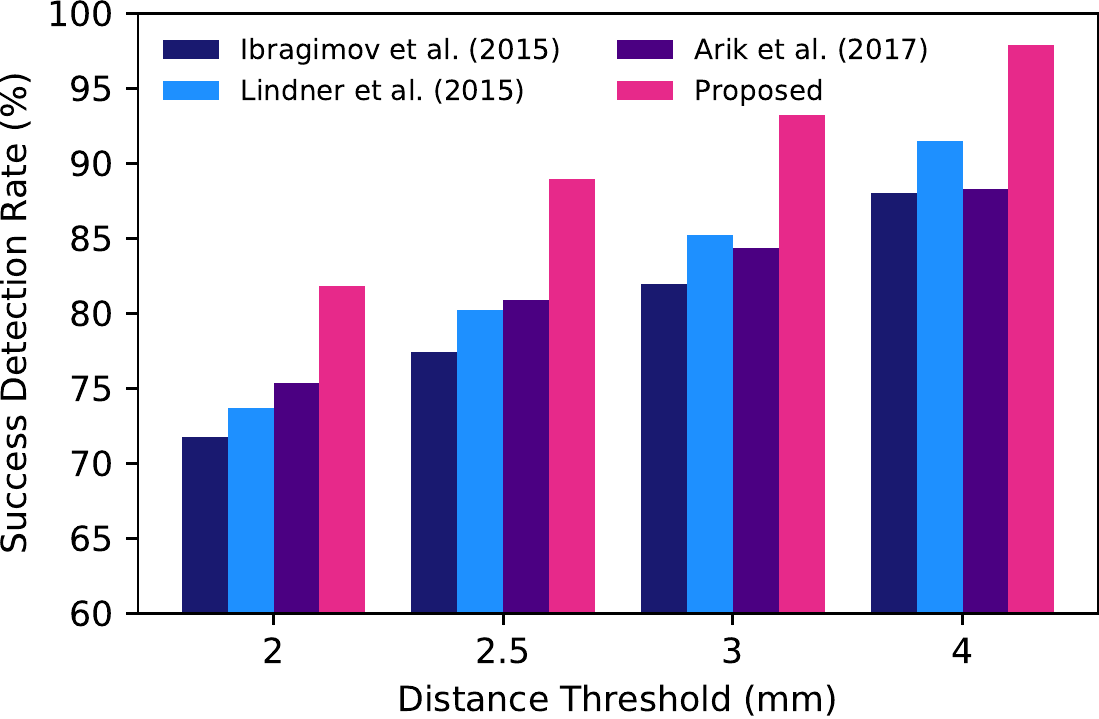}\\
			\small (a) Test1 
		\end{tabular}
		\begin{tabular}{@{}c@{}}
			\includegraphics[width=0.32\textwidth]{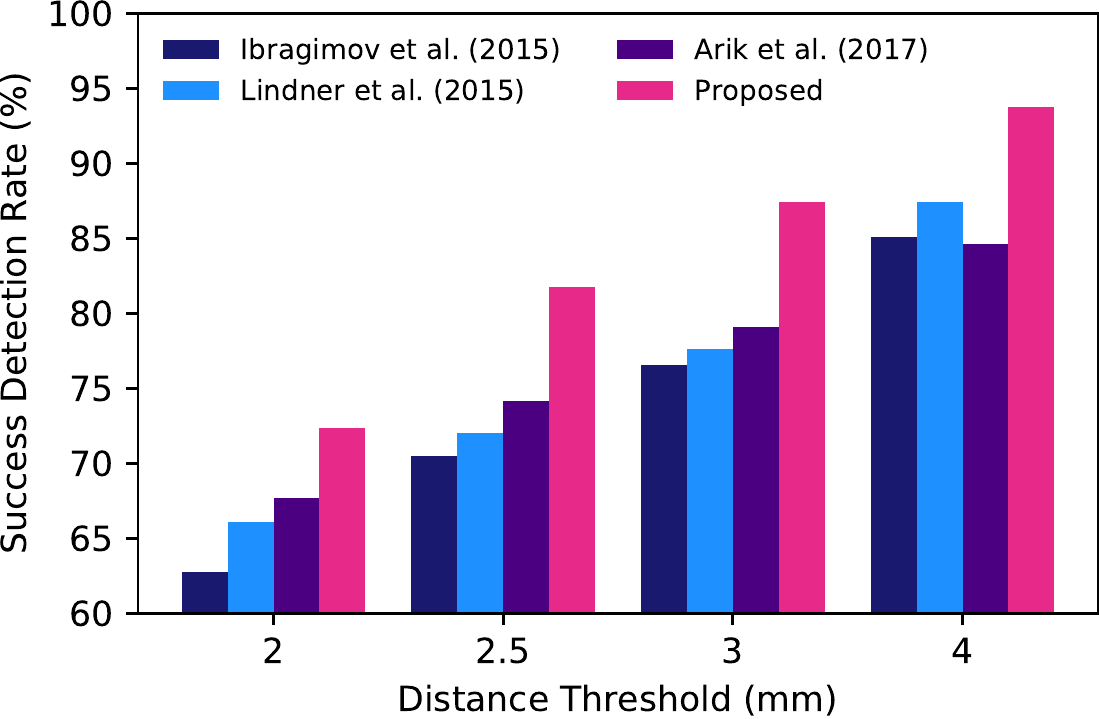}\\
			\small (b) Test2
		\end{tabular}
		\begin{tabular}{@{}c@{}}
			\includegraphics[width=0.32\textwidth]{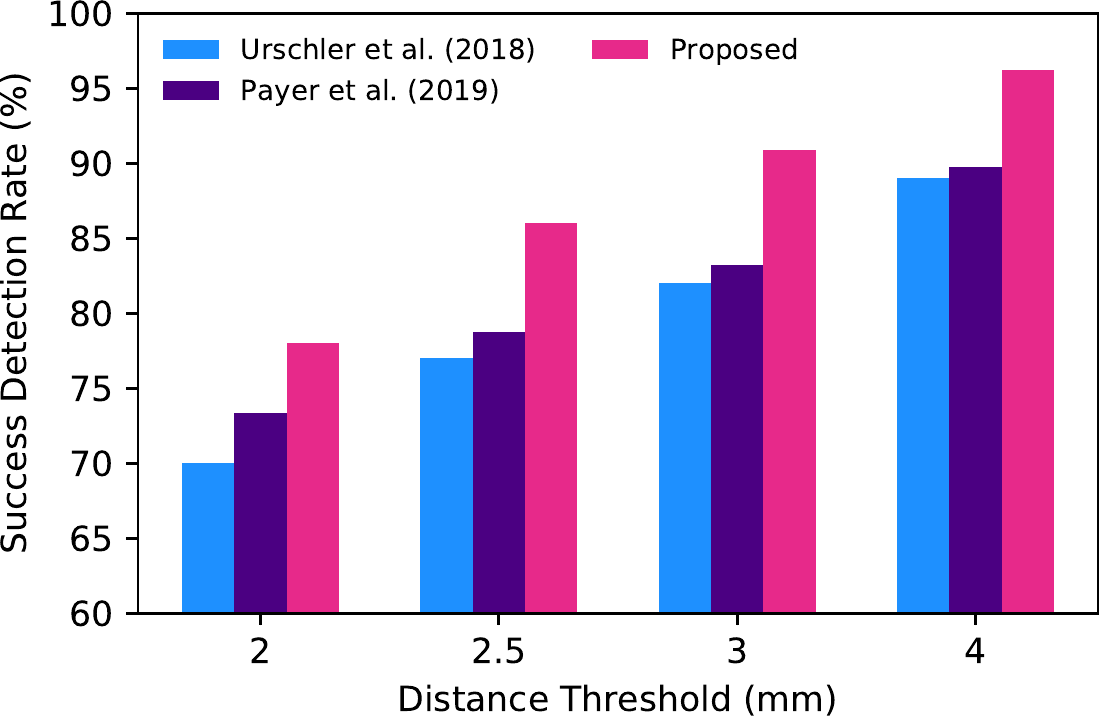}\\
			\small (c) Test1 and Test2
		\end{tabular}
		
		\caption{Success detection rates (SDRs) for landmark localization in cephalometric X-rays. SDRs obtained with the proposed method (Proposed) are shown together with SDRs reported in previous studies. Results are shown as \% over all landmarks and are given for four distance thresholds (2, 2.5, 3, 4\,mm). Results are shown separately for the two test sets, (a) Test1 containing 150 images, and (b) Test2 containing 100 images, (c) as well as for both test sets combined.}
		\label{fig:ceph_comparison}	
	\end{figure*}}
	
	\subsection{Ablation Study}
		To investigate whether the application of classification or the log-transform during training is truly beneficial for accurate landmark localization, we performed an ablation study with CCTA scans only, assuming results generalize to other datasets. For this, four additional networks for global multi-landmark localization were trained. These networks were trained with or without applying the log-transform with regression, and with or without using the classification output layer. For the classification-only network, a final landmark location was obtained by computing a weighted average of all predicted landmark locations. To obtain the final landmark location, the centers of the analyzed patches served as predicted landmark locations, while the posterior classification probabilities were used as weights during averaging. 
		
		Table~\ref{tab:ccta_results} shows that the proposed approach utilizing joint classification and regression of log-transformed displacement vectors achieved best performance (R\textsubscript{log} + C). Regression-only networks obtained smaller distance errors compared to classification-only networks. The addition of classification improved both regression-only networks, one using log-transform and one without it. The log-transform improved localization performance in the networks performing regression, with and without classification. When the approach for global localization utilizing joint classification and regression of log-transformed displacement vectors is combined with local single landmark localization (Table~\ref{tab:ccta_results}, Proposed ML), smaller distance errors were obtained compared to utilizing only global localization. Fig.~\ref{fig:CCTA_sdr_method} shows the obtained SDRs. Better SDRs were obtained by networks performing joint classification and regression compared to regression-only \REV{and classification-only} networks. However, the best results were obtained when joint classification and regression of log-transformed displacement vectors were used with a \REVV{global-to-local approach.}
	
	\subsection{Single Landmark Localization}
		\REVV{The proposed method employing joint classification and regression of log-transformed displacement vectors was evaluated for single landmark localization in CCTA by training one network for each of the eight cardiac landmarks. Table~\ref{tab:ccta_results} lists the obtained Euclidean distance errors (Proposed SL). With the exception of localization of the LRC and LH, the network trained for multi-landmark localization outperformed networks trained for single landmark localization. However, differences in performance were only significant for localization of the NCH and RH. Fig. \ref{fig:CCTA_sdr_method} shows the obtained SDRs. For a distance threshold of 0.5 mm, the SDR obtained with networks trained for single landmark localization was slightly better than the SDR obtained with a network trained for multi-landmark localization. However, this difference was only 0.6\%. For all other distance thresholds, better SDRs were obtained by the network trained for multi-landmark localization compared to networks trained for single landmark localization.}
		
	\subsection{Comparison with State-of-the-art}
		A number of methods have previously been proposed to localize anatomical landmarks in medical images.
    
	\subsubsection{Coronary CT Angiography}
	\label{sec:ccta_comparison}
	\begin{table*}
	\setlength{\tabcolsep}{3.5pt}
		\renewcommand{\arraystretch}{1.3}
		\centering
		\caption{\REV{Median (IQR) Euclidean distance errors (mm) between computed and reference landmark locations for eight landmarks in CCTA scans. Landmarks were automatically localized with the methods by Alansary et al. \cite{alansary2019evaluating}, Payer et al. \cite{payer2019integrating}, and with the proposed method (Proposed). Methods localize either single landmarks (SL) or multiple landmarks simultaneously (ML). Results are listed per landmark (NCRC, NCLC, LRC, LH, NCH, RH, RO, and LO) and for all landmarks together (All).}}
        \begin{tabular}{ll|cccccccc|c}
            Method  &    & NCRC        & NCLC        & LRC         & LH          & NCH         & RH          & RO          & LO          & All         \\ \hline
            Alansary et al. \cite{alansary2019evaluating}&SL & 3.35 (2.38)\ssymbol{3} & 3.35 (2.38)\ssymbol{3} & 3.35 (2.62)\ssymbol{3} & 3.35 (1.55)\ssymbol{3} & 3.67 (2.38)\ssymbol{3} & 3.67 (2.03)\ssymbol{3} & 3.35 (2.14)\ssymbol{3} & 2.60 (2.18)\ssymbol{3} & 3.35 (2.38)\ssymbol{3} \\
            Payer et al. \cite{payer2019integrating} & SL & 2.30 (2.03)\ssymbol{3} & 2.70 (1.89)\ssymbol{3} & 3.06 (2.85)\ssymbol{3} & 2.45 (1.80)* & 2.72 (1.57)  & 2.82 (1.50) & 2.03 (1.51)\ssymbol{3} & 2.69 (2.32)\ssymbol{3} & 2.55 (1.90)\ssymbol{3}\\\smallskip
            \REVV{Proposed} &  \REVV{SL} &  \REVV{1.99 (1.82)} &  \REVV{1.94 (1.32)} &  \REVV{1.69 (1.73)} &  \REVV{2.29 (1.46)} &  \REVV{2.68 (1.75)} &  \REVV{3.09 (1.82)} &  \REVV{1.48 (0.98)} &  \REVV{1.55 (1.02)} &  \REVV{2.03 (1.79)}\\
            Payer et al. \cite{payer2019integrating}& ML &
            1.79 (1.39) & 1.77 (1.52) & 1.90 (1.80) & 2.12 (1.39) & 2.50 (1.68) & 2.28 (1.36) & 1.30 (1.01) & 1.59 (1.17) & 1.90 (1.54)\\
            \REVV{Proposed}& \REVV{ML} & \REVV{1.85 (1.96)} & \REVV{1.80 (1.59} & \REVV{1.76 (1.67)} & \REVV{2.40 (1.58)} & \REVV{2.48 (1.72)} & \REVV{2.23 (1.42)} & \REVV{1.45 (1.20)} & \REVV{1.55 (0.97)} &	\REVV{1.87 (1.67)}\\
        \end{tabular}
        \begin{tablenotes}
         \small
         \item Significance outcome by the Wilcoxon Signed Rank test compared to the proposed method is indicated with * for p<0.05, $\dagger$ for p<0.01, and $\ddagger$ for p<0.001. Comparisons are made between methods performing single landmark localization (SL) or methods performing multi-landmark localization (ML).
         \end{tablenotes}
        \label{tab:ccta_PA}
    \end{table*}
		Previous methods have been proposed to specifically localize the aortic valve hinges, the aortic valve commissures, and the coronary ostia in cardiac CT scans. 
		
		Waecher et al. \cite{wachter2010patient} used pattern matching and reported distance errors of 1.0~$\pm$~0.8\,mm and 1.2~$\pm$~0.6\,mm for the right and left ostium, respectively. Wolterink et al. \cite{wolterink2019coronary} used a CNN to localize the ostia and obtained a mean distance error of 1.8~$\pm$~1.0\,mm. However, both methods were tested on small sets containing only 20\cite{wachter2010patient} or 36\cite{wolterink2019coronary} scans that might not have contained the anatomical deviation which was present in our test set comprising of 200 CCTA scans. Removing eight scans from our test set that show severe anatomical deviation (the right ostium located on the left side, the left ostium located on the right side), or cases where a stent is present in the pulmonary arteries, improves results for localization of the coronary ostia from \REVV{2.03~$\pm$~3.05\,mm to 1.75~$\pm$~1.84\,mm}.

		For the aortic valve commissures and the aortic valve hinges, the proposed method obtained mean distance errors of \REVV{2.33~$\pm$~1.90\,mm and 2.61~$\pm$~1.44\,mm,} respectively. Zheng et al. \cite{zheng2012automatic} exploited voxel classification with landmark specific probabilistic boosting trees and reported mean distance errors of 2.17~$\pm$~1.31\,mm, 2.09~$\pm$~1.18\,mm, and 2.07~$\pm$~1.53\,mm for the aortic valve commissures, the aortic valve hinges, and the coronary ostia, respectively. Landmarks were localized in in C-arm CT scans with a voxel size ranging between 0.70 and 0.84 mm. 
	
		Elattar et al. \cite{elattar2016automatic} applied a local rule-based approach and combined results obtained for localization of the aortic valve hinges and the coronary ostia in 40 CCTA scans with a voxel size varying from 0.44 to 0.9\,mm. The analysis led to a mean distance error of 2.81~$\pm$~2.08\,mm. Al et al. \cite{al2018automatic} also combined results obtained for localization of all eight landmarks in 71 CCTA scans using cross-validation and obtained a mean distance error of 2.04~$\pm$~1.11\,mm. Voxel sizes of used CCTA scans were not reported. Computing the same measure, the mean distance error obtained for localization of the eight cardiac landmarks, we obtained a distance error of \REVV{2.36~$\pm$~1.24\,mm}. 
		
		These aforementioned methods have been developed and tested on different CCTA datasets than used in our work. Hence, a comparison between the results should only be used as an indication of the performance. To enable a direct comparison of our methods with previous work, we have tested the very recently proposed methods by Alansary et al. \cite{alansary2019evaluating}, who employ reinforcement learning and localize a single landmark at the time, and Payer et al. \cite{payer2019integrating}, who employ heatmap regression to localize either a single landmark or multiple landmarks jointly, on our data. For this, we used code made publicly available by the authors\footnote{https://github.com/amiralansary}\footnote{https://www.github.com/christianpayer}. The results are listed in Table~\ref{tab:ccta_PA}. The Wilcoxon Signed Rank test was used to test for significance. Results show that our method performing single landmark localization significantly outperformed the method proposed by Alansary et al. \cite{alansary2019evaluating}. Furthermore, we significantly outperform the method proposed by \REVV{Payer et al. \cite{payer2019integrating} when trained for single landmark localization. When comparing our method for multi-landmark localization with the multi-landmark localization proposed by Payer et al. \cite{payer2019integrating}, differences in performance are not significant.}
		
		On average, the processing time per scan was 0.29~$\pm$~0.56 seconds for the method proposed by Alansary et al. \cite{alansary2019evaluating}, and 0.42~$\pm$~0.13 seconds and 0.49~$\pm$~0.28 seconds for the method proposed by Payer et al. \cite{payer2019integrating} for single landmark localization and multi-landmark localization, respectively.
		\REVV{On average, the processing times per scan for our method were 0.04~$\pm$~0.01 and 0.06~$\pm$~0.05 seconds for single landmark localization and multi-landmark localization, respectively.}
		
    \subsubsection{Olfactory MR}
        To the best of our knowledge, no landmark localization methods have been evaluated for localization of the olfactory bulbs in MRI. To compare our proposed method with state-of-the-art landmark localization methods, we have evaluated the publicly available methods by Alansary et al. \cite{alansary2019evaluating} and Payer et al. \cite{payer2019integrating} as for the landmarks in CCTA (see section \ref{sec:ccta_comparison}). Results are listed in Table~\ref{tab:bulb_PA}. Comparing our method for localization of one olfactory bulb per scan, we significantly outperform other methods performing single landmark localization. The difference between methods performing localization of both olfactory bulbs simultaneously was not significant. On average, the processing time per scan was 0.78~$\pm$~1.53 seconds for the method proposed by Alansary et al. \cite{alansary2019evaluating}, and 0.38~$\pm$~0.003 seconds and 0.44~$\pm$~0.25 seconds for the method proposed by Payer et al. \cite{payer2019integrating} for single landmark localization and multi-landmark localization, respectively. For the proposed method, the processing time per scan for single landmark localization and multi-landmark localization were 0.07~$\pm$~0.008 and 0.07~$\pm$~0.01 seconds per scan, respectively.
        
        \begin{table}
    		\renewcommand{\arraystretch}{1.3}
    		\centering
    		\caption{\REV{Median (IQR) Euclidean distance errors (mm) between computed and reference landmark locations for the right and left olfactory bulb in MRI. Landmarks were automatically localized with the methods by Alansary et al. \cite{alansary2019evaluating}, Payer et al. \cite{payer2019integrating}, and the proposed method (Proposed). Methods localize either single landmarks (SL) or multiple landmarks simultaneously (ML). Results are listed per landmark (Right Bulb, Left Bulb) and for both bulbs together (Both Bulbs).}}
            \begin{tabular}{ll|cc|c}
                Method &  &Right Bulb  & Left Bulb  & Both Bulbs  \\\hline
                Alansary et al.\cite{alansary2019evaluating}& SL & 1.28 (1.46) & 1.44 (1.42)* & 1.41 (1.42)\ssymbol{2} \\
                Payer et al. \cite{payer2019integrating}& SL & 1.67 (1.48)* &  1.38 (2.22) &  1.55 (2.00)\ssymbol{2} \\\smallskip
                \REVV{Proposed} & \REVV{SL} & \REVV{0.93 (0.74)} & \REVV{0.99 (0.94)} & \REVV{0.95 (0.94)} \\
                Payer et al. \cite{payer2019integrating}& ML & 0.91 (1.06) & 0.76 (1.01) & 0.89 (0.97) \\
                \REVV{Proposed} & \REVV{ML} & \REVV{0.87 (1.36)} & \REVV{0.90 (0.58)} & \REVV{0.90 (0.85)} \\
            \end{tabular}
            \begin{tablenotes}
         \small
         \item Significance outcome by the Wilcoxon Signed Rank test compared to the proposed method is indicated with * for p<0.05 and $\dagger$ for p<0.01. Comparisons are made between methods performing single landmark localization (SL) or methods performing multi-landmark localization (ML).
         \end{tablenotes}
            \label{tab:bulb_PA}
        \end{table}
            	
	\subsubsection{Cephalometric X-rays}
        Previous methods have been proposed to localize landmarks in cephalometric
        X-rays. Ibragimov et al. \cite{ibragimov2015computerized}, Lindner et al.
        \cite{lindner2015fully}, and Urschler et al.\cite{urschler2018integrating} all employed conventional machine learning, while Arik et al. \cite{arik2017fully} and Payer et al. \cite{payer2019integrating} both proposed a CNN to localize landmarks in cephalometric X-rays. 
        
        Fig.~\ref{fig:ceph_comparison} shows a comparison between the SDRs obtained in previous studies and the SDRs obtained with the proposed method. Payer et al. \cite{payer2019integrating} reported the percentage of outliers. Hence, for comparison with our results, we reformulated their results into SDRs. For all distance thresholds, our method \REVV{obtained better SDRs compared to those obtained in previous studies}.
        
        Reported processing times for the method proposed by Lindner et al.
        \cite{lindner2015fully}, Urschler et al.\cite{urschler2018integrating}, and Payer et al. \cite{payer2019integrating} were 5, 56, and 2 seconds per scan, respectively. However, for the proposed method, the processing time for localization of all landmarks was on average 0.05~$\pm$~0.009 seconds per scan. 
	
	\section{Discussion}
	An automatic method for anatomical landmark localization in medical images has been proposed. The method \REVV{employs global-to-local analysis where initially a fully convolutional neural network predicts the locations of multiple landmarks simultaneously. Subsequently, specialized FCNNs refine the global landmark locations. For global multi-landmark localization, an FCNN analyzes 2D or 3D images of arbitrary size in a patch-based manner. For every patch in an image, regression is used to predict displacement vectors that point from the center of the patch to landmarks of interest. Simultaneously, classification is used to predict the presence of landmarks of interest in each image patch. The global landmark locations are obtained by a weighted average of the displacement vectors predicted by regression, using posterior probabilities predicted by classification as weights. Subsequently, specialized FCNNs refine the global landmark locations by analyzing a local sub-image around each landmark in a similar manner, performing regression and classification simultaneously and combining the results.}
	
	The method was evaluated using three different datasets, namely 3D CCTA scans, 3D olfactory MR scans and 2D cephalometric X-rays. Results demonstrate that the method is able to localize landmarks with high accuracy in medical images differing in modality, dimensionality and depicted anatomy. Results obtained for the localization of the aortic valve commissures, the aortic valve hinges, and the coronary ostia in CCTA are comparable with the intra-observer variability and second observer performance. Previous methods analyzed images at higher resolution and reported slightly better results. However, these methods have been developed and tested on different CCTA datasets than used in our work and therefore, a comparison between the results should only be used as an indication of the performance. \REV{Ideally, landmarks would be localized in the native resolution. However, due to hardware limitation, we have resampled the scans prior to analysis to reduce the image resolution. It is worth noting that memory limitations prohibited analysis of complete images at the native resolution. Hence, hardware limitations require partitioning of images during inference for analysis at a resolution close to the native resolution. Our preliminary experiments showed that increasing the image resolution had a negative impact on the performance when analyzing complete images. However, addressing }the hardware limitations and analyzing scans with a higher resolution will probably lead to lower distance errors in mm. 
	
	Using the CCTA dataset, we have shown that joint regression and classification improves upon \REV{classification-only}. \REV{For the classification-only networks, landmarks were localized by computing a weighted average of the predicted landmark locations. For this, the center of analyzed patches were considered the predicted landmark locations while the classification output, i.e. posterior probabilities, served as weights during averaging. A different approach for the final decision could also be considered. For example, only the center of the patch with the highest classification probability could have been taken into account or only patches with a posterior probability higher than a threshold could have been used. However, the ability of the classification-only network to precisely localize a landmark will always be limited by its patch-size. A deeper neural network that could perform voxel-based classification could therefore be more precise compared to a patch-based classification network. Nevertheless, a voxel-based classification network would demand balancing of the data during training due to a high class imbalance between landmark and background voxels. Furthermore, voxel-based analysis is more computationally demanding compared to the here proposed patch-based classification network. Since landmark localization is typically a prerequisite for subsequent, more complex medical image analysis \cite{rohr2001landmark, miao2012automatic, murphy2011semi, wang2018fast, han_robust_2014, alam2018medical, oktay2017stratified, wang2016benchmark, al2018automatic, torosdagli2018deep, kasel2013standardized, ionasec2008dynamic, zhengautomatic2010}, localization speed may be important.}
	
	Additionally, we have also shown that joint regression and classification improves upon regression-only landmark localization. For regression-only methods, the final landmark location was predicted by computing the average over all landmark locations obtained with the predicted displacement vectors. Hence, independent of their distance to the landmark, all patches contributed equally. Inspection of the results showed that predictions from patches far from the landmark of interest resulted in larger distance errors compared to those made from patches close to the landmark of interest. Hence, equally weighting all predictions resulted in larger distance errors compared to joint classification and regression. With joint regression and classification, such errors were mitigated by weighting the displacement vectors using the posterior probabilities obtained from classification. Namely, patches farther from the landmark of interest received lower posterior probabilities, thereby reducing the influence on the final landmark prediction. Employing a log-transform for displacement further improved localization. This is likely caused by the nature of the log-transform under the influence of the mean absolute error loss during training; i.e. during training, prediction errors from patches close to the landmark of interest are more heavily penalized than predictions from patches far from the landmark of interest.
	
	\REVV{Networks trained for multi-landmark localization obtained slightly better results compared to networks trained for single landmark localization in CCTA. However, no statistically significant difference between the two approaches was found, indicating that our proposed method could be used for single- as well as for multi-landmark localization.} 
	
	Visual inspection of the results obtained with our proposed method on the CCTA dataset showed that larger Euclidean distance errors were obtained in images in which anatomical abnormalities were present, such as the right ostium being located on the left side of the patient. Training the network with more images that depict these types of anatomical deviation or modeling of these anatomical deviations by exploiting data augmentation, such as elastic transformations of images, could be beneficial to increase the variation in the dataset and ultimately improve localization.
	
	In contrast to previous work that required segmentation of the aorta \cite{elattar2016automatic, zheng2012automatic, wachter2010patient}, the here proposed method does not require any prior segmentation steps. Moreover, during inference, the method analyzes complete images and thus it is capable of localizing target landmarks in large 3D image volumes with high speed. As demonstrated by the results, direct learning from the data without preprocessing steps incorporating knowledge about the anatomy leads to accurate localization of the eight cardiac landmarks. \REVV{When comparing with recent methods \cite{alansary2019evaluating, payer2019integrating}, our method outperforms results obtained by Alansary et al. \cite{alansary2019evaluating} and Payer et al. \cite{payer2019integrating} in the localization of single landmarks. Furthermore, our method performs on par compared to the method proposed by Payer et al. \cite{payer2019integrating} for multi-landmark localization.} 
	
	\REVV{Contrary to earlier approaches, the here proposed method can be used for both single and multi-landmark localization. Furthermore, our method is able to localize landmarks faster compared to competing methods. For pre-operative applications or offline tasks, such as the initialization of segmentation methods \cite{oktay2017stratified}, localization speed might be less important but for real-time applications, such as intra-operative applications, speed may be crucial \cite{alam2018medical}.} For landmark localization in both test sets of the cephalometric X-ray challenge, our proposed method outperformed previous methods, having an error close to the variability between the two observers.
	
	\section{Conclusion}
	We have shown that the proposed method is able to localize landmarks in 2D and 3D medical images of arbitrary size, acquired with three different imaging modalities and depicting different anatomical coverage. \REVV{The method localizes multiple or single landmarks} with high accuracy and speed, making it suitable for application in studies including a large number of images or real-time localization.

	\section*{Acknowledgments}
	This work is part of the research program Deep Learning for Medical Image Analysis with project number P15-26, financed by the Dutch Technology Foundation with contribution by Philips Healthcare.
	
	\IEEEtriggeratref{37}
	\bibliography{bibLMD}{}
	\bibliographystyle{IEEEtran}
\end{document}